\RequirePackage{fix-cm}
\documentclass[smallextended]{svjour3}       
\smartqed  
\usepackage{graphicx}
\usepackage{amsfonts,amsmath,amssymb,mathrsfs}
\usepackage[usenames,dvipsnames]{xcolor}
\usepackage{color}

\def\be{\begin{equation}}
\def\ee{\end{equation}}
\def\beq{\begin{eqnarray}}
\def\eeq{\end{eqnarray}}

\def\nn{\nonumber}

\begin{document}

\title{Astrophysical black holes as natural laboratories\\for fundamental physics and strong-field gravity}

\titlerunning{Astrophysical black holes}        

\author{Emanuele Berti}


\institute{E. Berti \at
           Department of Physics and Astronomy, The University of
           Mississippi, University, MS 38677, USA\\
           California Institute of Technology, 
           Pasadena, CA 91109, USA\\
              \email{berti@phy.olemiss.edu}          
}

\date{Received: date / Accepted: date}

\maketitle

\begin{abstract}
Astrophysical tests of general relativity belong to two categories: 1)
``internal'', i.e. consistency tests {\em within} the theory (for
example, tests that astrophysical black holes are indeed described by
the Kerr solution and its perturbations), or 2) ``external'',
i.e. tests of the many proposed extensions of the theory. I review
some ways in which astrophysical black holes can be used as natural
laboratories for both ``internal'' and ``external'' tests of general
relativity. The examples provided here (ringdown tests of the black
hole ``no-hair'' theorem, bosonic superradiant instabilities in
rotating black holes and gravitational-wave tests of massive
scalar-tensor theories) are shamelessly biased towards recent research
by myself and my collaborators. Hopefully this colloquial introduction
aimed mainly at astrophysicists will convince skeptics (if there are
any) that space-based detectors will be crucial to study fundamental
physics through gravitational-wave observations.  
\end{abstract}

\section{Introduction}

The foundations of Einstein's general relativity (GR) are very well
tested in the regime of weak gravitational fields, small spacetime
curvature and small velocities \cite{Will:2005va}. It is generally
believed, on both theoretical and observational grounds (the most
notable observational motivation being the dark energy problem), that
Einstein's theory will require some modification or extension at high
energies and strong gravitational fields, and these modifications
generally require the introduction of additional degrees of freedom in
the theory \cite{Clifton:2011jh}.

Because GR is compatible with all observational tests in weak-gravity
conditions, a major goal of present and future experiments is to probe
astrophysical systems where gravity is, in some sense, strong. The
strength of gravity can be measured either in terms of the
gravitational field $\varphi\sim M/r$, where $M$ is the mass and $r$
the size of the system in question\footnote{Throughout this paper I
  will use geometrical units $G=c=1$.}, or in terms of the curvature
\cite{Psaltis:2008bb}. A quantitative measure of curvature are tidal
forces, related to the components $R^r_{~0r0}\sim M/r^3$ of the
Riemann tensor associated to the spacetime metric $g_{ab}$
\cite{Misner:1974qy}. The field strength is related to typical
velocities of the system by the virial theorem ($v\sim \varphi^{1/2}
\sim \sqrt{M/r}$) so it is essentially equivalent to the
post-Newtonian small velocity parameter $v$ (or $v/c$ in ``standard''
units). One could argue that ``strong curvature'' is in some ways more
fundamental than ``strong field'', because Einstein's equations relate
the stress-energy content of the spacetime to its curvature (so that
``curvature is energy'') and because the curvature (not the field
strength) enters the Lagrangian density in the action principle
defining the theory: cf. e.g. Eq.~(\ref{lagrangianGB}) below.

It is perhaps underappreciated that in astrophysical systems one can
``probe strong gravity'' by observations of weak gravitational fields,
and vice versa, observations in the strong-field regime may not be
able to tell the difference between GR and its alternatives or
extensions.

The possibility to probe strong-field effects using weak-field binary
dynamics is nicely illustrated by the ``spontaneous scalarization''
phenomenon discovered by Damour and Esposito-Far\'ese
\cite{Damour:1993hw}. The idea is that the coupling of the scalar with
matter can allow some scalar-tensor theories to pass all weak-field
tests, while at the same time introducing macroscopically (and
observationally) significant modifications in the structure of neutron
stars (NSs). If spontaneous scalarization occurs\footnote{The
  astrophysical plausibility of spontaneous scalarization is supported
  by detailed studies of stellar structure
  \cite{Damour:1996ke,Salgado:1998sg}, numerical simulations of
  collapse \cite{Shibata:1994qd,Harada:1996wt,Novak:1997hw} and
  stability analyses \cite{Harada:1997mr}. While the strength of
  spontaneous scalarization phenomena is already strongly constrained
  by observations of binary pulsars \cite{Freire:2012mg},
  semiclassical vacuum instabilities seem to offer a viable mechanism
  to ``seed'' nonzero scalar fields in stars
  \cite{Lima:2010na,Pani:2010vc}.}, the masses of the two stars in a
binary can in principle be very different from their GR
values. Therefore the dynamics of NS binaries will be significantly
modified even when the binary members are sufficiently far apart that
$v\sim \sqrt{M/r}\ll 1$. For this reason, ``weak-field'' observations
of binary pulsars can strongly constrain a strong-field phenomenon
such as spontaneous scalarization \cite{Freire:2012mg}.

On the other hand, measurements of gas or particle dynamics in
strong-field regions around the ``extremely relativistic'' Kerr black
hole (BH) spacetime are {\em not} necessarily smoking guns of
hypothetical modifications to general relativity.
The reason is that classic theorems in Brans-Dicke theory
\cite{1971ApJ...166L..35T,Hawking:1972qk,Bekenstein:1996pn}, recently
extended to generic scalar-tensor theories and $f(R)$ theories
\cite{Sotiriou:2008rp,Psaltis:2007cw}, show that solutions of the
field equations in vacuum always include the Kerr metric as a special
case. The main reason is that many generalizations of GR admit the
vacuum equations of GR itself as a special case. This conclusion may
be violated e.g. in the presence of time-varying boundary conditions,
that could produce ``BH hair growth'' on cosmological timescales
\cite{Horbatsch:2011ye} and dynamical horizons \cite{Faraoni:2012sf}.

The Kerr solution is so ubiquitous that probes of the Kerr {\em
  metric} alone will not tell us whether the correct theory of gravity
is indeed GR. However, the {\em dynamics} of BHs (as manifested in
their behavior when they merge or are perturbed by external agents
\cite{Barausse:2008xv}) will be very different in GR and in
alternative theories. In this sense, gravitational radiation (which
bears the imprint of the dynamics of the gravitational field) has the
potential to tell GR from its alternatives or extensions.

To wrap up this introduction: our best bet to probe strong-field
dynamics are certainly BHs and NSs, astronomical objects for which
both $\varphi\sim M/r$ and the curvature $\sim M/r^3$ are
large. However: 1) there is the definite possibility that weak-field
observations may probe strong gravity, as illustrated e.g. by the
spontaneous scalarization phenomenon; and 2) measurements of the
metric around BH spacetimes will not be sufficient to probe GR, but
{\em dynamical} measurements of binary inspiral and merger dynamics
will be sensitive to the dynamics of the theory.

\section{Finding contenders to general relativity}

Let us focus for the moment on ``external'' tests, i.e. test of GR
versus alternative theories of gravity. What extensions of GR can be
considered serious contenders? A ``serious'' contender (in this
author's opinion) should at the very least be well defined in a
mathematical sense, e.g. by having a well posed initial-value
problem. From a phenomenological point of view, the theory must also
be simple enough to make physical predictions that can be validated by
experiments (it is perhaps a sad reflection on the current state of
theoretical physics that one should make such a requirement
explicit!).

An elegant and comprehensive overview of theories that have been
studied in the context of space-based gravitational-wave (GW)
astronomy is presented in~\cite{Gair:2012nm}. Here I focus on a
special subclass of extensions of GR whose implications in the context
of Solar-System tests, stellar structure and GW astronomy have been
explored in some detail. I will give a ``minimal'' discussion of these
theories, with the main goal of justifying the choice of massive
scalar-tensor theories as a particularly simple and interesting
phenomenological playground.

Among the several proposed extensions of GR (see
e.g.~\cite{Clifton:2011jh} for an excellent review), theories that can
be summarized via the Lagrangian density
\begin{eqnarray}
\label{lagrangianGB}
{\cal L}&&= f_0(\phi)\color{black}R \\
&&-\varpi(\phi) g^{ab}\partial_a\phi \partial_b\phi-M(\phi)+
{\cal L}_{\rm mat}\left[\Psi,A^2(\phi)g_{ab}\right]\,\nn\\
&&+f_1(\phi) {\cal R}^2_{\rm GB}
+ f_2(\phi)R_{abcd}{}^*R^{abcd}\,\nn
\end{eqnarray}
have rather well understood observational implications for cosmology,
Solar System experiments, the structure of compact stars and
gravitational radiation from binary systems.

In the Lagrangian given above $\phi$ is a scalar-field degree of
freedom (not to be confused with the gravitational field strength
$\varphi$ introduced earlier); $R_{abcd}$ is the Riemann tensor,
$R_{ab}$ the Ricci tensor and $R$ the Ricci scalar corresponding to
the metric $g_{ab}$; $\Psi$ denotes additional matter fields. The
functions $f_i(\phi)$ ($i=0\,,1\,,2$), $M(\phi)$ and $A(\phi)$ are in
principle arbitrary, but they are not all independent. For example,
field redefinitions allow us to set either $f_0(\phi)=1$ or
$A(\phi)=1$, which corresponds to working in the so-called
``Einstein'' or ``Jordan'' frames, respectively. This Lagrangian
encompasses models in which gravity is coupled to a single scalar
field $\phi$ in all possible ways, including all linearly independent
quadratic curvature corrections to GR.

Scalar-tensor gravity with generic coupling, sometimes called
Bergmann-Wagoner theory \cite{Bergmann:1968ve,Wagoner:1970vr},
corresponds to setting $f_1(\phi)=f_2(\phi)=0$ in
Eq.~(\ref{lagrangianGB}). This is one of the oldest and best-studied
modifications of GR. If we further specialize to the case where
$A(\phi)=1$, $f_0(\phi)=\phi$, $\varpi(\phi)=\omega_{\rm BD}/\phi$ and
$M(\phi)=0$ we recover the ``standard'' Brans-Dicke theory of gravity
in the Jordan frame \cite{Brans:1961sx}; the Einstein frame
corresponds to setting $f_0(\phi)=1$ instead. In a Taylor expansion of
$M(\phi)$, the term quadratic in $\phi$ introduces a nonzero mass for
the scalar (see e.g. \cite{Alsing:2011er}). GR is recovered in the
limit $\omega_{\rm BD}\to \infty$.

Initially motivated by attempts to incorporate Mach's principle into
GR, scalar-tensor theories have remained popular both because of their
relative simplicity, and because scalar fields are the simplest
prototype of the additional degrees of freedom predicted by most
unification attempts \cite{Fujii:2003pa}. Bergmann-Wagoner theories
are less well studied than one might expect, given their long
history\footnote{Tensor multi-scalar theories of gravity have also
  been investigated in depth (see e.g. \cite{Damour:1992we} and
  references therein), but we will not consider them here.}. These
theories can be seen as the low-energy limit of several proposed
attempts to unify gravity with the other interactions or, more
pragmatically, as mathematically consistent alternatives to GR that
can be used to understand which features of the theory are
well-tested, and which features need to be tested in more detail
\cite{EspositoFarese:2004cc}.
Most importantly, they meet all of the basic requirements of
``serious'' contenders to GR, as defined above.  They are well-posed
and amenable to numerical evolutions \cite{Salgado:2008xh}, and in
fact numerical evolutions of binary mergers in scalar-tensor theories
have already been performed for both BH-BH \cite{Healy:2011ef} and
NS-NS \cite{Barausse:2012da} binaries. At present, the most stringent
bound on the coupling parameter of standard Brans-Dicke theory
($\omega_{\rm BD}>40,000$) comes from Cassini measurements of the
Shapiro time delay \cite{Will:2005va}, but binary pulsar data are
rapidly becoming competitive with the Cassini bound: observations of
binary systems containing at least one pulsar, such as the
pulsar-white dwarf binary PSR J1738+0333, already provide very
stringent bounds on Bergmann-Wagoner theories \cite{Freire:2012mg}.

The third line of the Lagrangian (\ref{lagrangianGB}) describes
theories quadratic in the curvature. The requirement that the field
equations should be of second order means that corrections quadratic
in the curvature must appear in the Gauss-Bonnet (GB) combination
${\cal R}^2_{\rm GB}=R^2-4R_{ab}R^{ab}+R_{abcd}R^{abcd}$.
We also allow for a dynamical Chern-Simons correction proportional to
the wedge product $R_{abcd}{}^*R^{abcd}$
\cite{Alexander:2009tp}. Following \cite{Pani:2011xm}, we will call
these models ``extended scalar-tensor theories''. These theories have
been extensively investigated from a phenomenological point of view:
the literature includes studies of Solar-system tests
\cite{Amendola:2007ni,Amendola:2008vd}, BH solutions and dynamics
\cite{Yunes:2009hc,Pani:2011gy,Yagi:2012ya,Motohashi:2011pw,Motohashi:2011ds},
NS structure \cite{Yunes:2009ch,AliHaimoud:2011fw,Pani:2011xm} and
binary dynamics \cite{Yagi:2011xp,Yagi:2012vf,Yagi:2013mbt}. While the
interest of this class of theories is undeniable, and recent work has
highlighted very interesting phenomenological consequences for the
dynamics of compact objects, it is presently unclear whether they
admit a well defined initial value problem and whether they are
amenable to numerical evolutions. In analytical treatments these
theories are generally regarded as ``effective'' rather than
fundamental (see e.g. \cite{Yagi:2011xp} for a discussion), and
treated in a small-coupling approximation that simplifies the field
equations and ensures that the field equations are of second order.

The Lagrangian (\ref{lagrangianGB}) is more generic than it may
seem. For example, it describes -- at least at the formal level --
theories that replace the Ricci scalar $R$ by a generic function
$f(R)$ in the Einstein-Hilbert action, because these theories can
always be cast as (rather anomalous) scalar-tensor theories via
appropriate variable redefinitions
\cite{Sotiriou:2008rp,DeFelice:2010aj}.  Unfortunately the mapping
between $f(R)$ theories and scalar-tensor theories is in general
multivalued, and one should be very careful when considering the
scalar-tensor ``equivalent'' of an $f(R)$ theory (see
e.g.~\cite{Jaime:2010kn}). Recently popular theories that are not
encompassed by the Lagrangian above include e.g. Einstein-aether
theory \cite{Jacobson:2008aj}, Ho$\check{\rm r}$ava gravity
\cite{Horava:2009uw}, Bekenstein's TeVeS \cite{Bekenstein:2004ne},
massive gravity theories \cite{deRham:2010kj} and ``Eddington inspired
gravity'' \cite{Banados:2010ix}, which is equivalent to GR in vacuum,
but differs from it in the coupling with matter.

An overview of these theories is clearly beyond the scope of this
paper. From now on I will focus on the surprisingly overlooked fact
that theories of the Bergmann-Wagoner type, which are among the
simplest options to modify GR, allow us to introduce very interesting
dynamics by simply giving a nonzero mass to the scalar field. Scalar
fields predicted in unification attempts are generally massive, so
this ``requirement'' is in fact very natural. I will now argue that
massive scalar fields give rise to extremely interesting phenomena in
BH physics (Section \ref{sec:BHs}) and binary dynamics (Section
\ref{sec:GWs}).

\section{Black hole dynamics and superradiance}
\label{sec:BHs}

With the caveat that measurements based on the Kerr {\em metric} alone
do not necessarily differentiate between GR and alternative theories
of gravity, BHs are ideal astrophysical laboratories for strong field
gravity. Recent results in numerical relativity (see
e.g. \cite{Buonanno:2006ui,Berti:2007fi}) confirmed that the dynamics
of BHs can be approximated surprisingly well using linear perturbation
theory (see Chandrasekhar's classic monograph \cite{chandrasekhar-83}
for a review).
In perturbation theory, the behavior of test fields of any spin
(e.g. $s=0\,,1\,,2$ for scalar, electromagnetic and gravitational
fields) can be described in terms of an effective potential
\cite{chandrasekhar-83,Berti:2009kk}. For massless scalar
perturbations of a Kerr BH, the potential is such that: 1) it goes to
zero at the BH horizon, which (introducing an appropriate radial
``tortoise coordinate'' $r_*$ \cite{chandrasekhar-83}) corresponds to
$r_*\to -\infty$; 2) it has a local maximum located (roughly) at the
light ring; 3) it tends to zero as $r_*\to \infty$. A nonzero scalar
mass does not qualitatively alter features 1) and 2), but it creates a
{\em nonzero potential barrier} such that $V\to m^2$ (where $m$ is the
mass of the field in natural units $G=c=\hbar=1$) at infinity. Because
of the nonzero potential barrier, the potential can now accommodate
{\em quasibound states} in the potential well located between the
light-ring maximum and the potential barrier at infinity
(cf. e.g. Fig.~7 of \cite{Arvanitaki:2010sy}). These states are
quasibound because the system is dissipative. In fact, under
appropriate conditions the system can actually be unstable. The stable
or unstable nature of BH perturbations is detemined by the shape of
the potential and by a well-known feature of rotating BHs: the
possibility of {\em superradiant amplification} of perturbation
modes. I will begin by discussing stable perturbations in Section
\ref{sec:stable}, and then I will turn to superradiantly unstable
configurations in Section \ref{sec:unstable}.

\subsection{Stable dynamics in GR: quasinormal modes}
\label{sec:stable}

Massless (scalar, electromagnetic or gravitational) perturbations of a
Kerr BH have a ``natural'' set of boundary conditions: we must impose
that waves can only be ingoing at the horizon (which is a one-way
membrane) and outgoing at infinity, where the observer is
located. Imposing these boundary conditions gives rise to an
eigenvalue problem with complex eigenfrequencies, that correspond to
the so-called BH quasinormal modes \cite{Berti:2009kk}. The nonzero
imaginary part of the modes is due to damping (radiation leaves the
system both at the horizon and at infinity), and its inverse
corresponds to the damping time of the perturbation. By analogy with
damped oscillations of a ringing bell, the gravitational radiation
emitted in these modes is often called ``ringdown''.

The direct detection of ringdown frequencies from perturbed BHs will
provide stringent {\em internal} tests that astrophysical BHs are
indeed described by the Kerr solution. The possibility to carry out
such a test depends on the signal-to-noise ratio (SNR) of the observed
GWs: typically, SNRs larger than $\sim 30$ should be sufficient to
test the Kerr nature of the remnant \cite{Berti:2005ys}.  While these
tests may be possible using Earth-based detectors, they will probably
require observations of relatively massive BH mergers with total mass
$\sim 10^2M_\odot$. A detection of such high-mass mergers would be a
great discovery in and by itself, given the dubious observational
evidence for intermediate-mass BHs \cite{Miller:2003sc}. On the other
hand, the existence of massive BHs with $M\gtrsim 10^5M_\odot$ is well
established, and space-based detectors such as (e)LISA
\cite{Danzmann:1998,AmaroSeoane:2012km,AmaroSeoane:2012je} have a
formidable potential for observing the mergers of the lightest
supermassive BHs with large SNR throughout the Universe (see
e.g. Fig.~16 of \cite{AmaroSeoane:2012km}).  Any such observation
would yield stringent ``internal'' strong-field tests of
GR. Furthermore, ringdown observations can be used to provide
extremely precise measurements of the remnant spins
\cite{Berti:2005ys}.  Since the statistical distribution of BH spins
encodes information on the past history of assembly and growth of the
massive BH population in the Universe \cite{Berti:2008af}, spin
measurements can be used to discriminate between astrophysical models
that make different assumptions on the birth and growth mechanism of
massive BHs \cite{Gair:2010bx,Sesana:2010wy}.

\begin{figure}[th]
\includegraphics[width=18pc,clip=true]{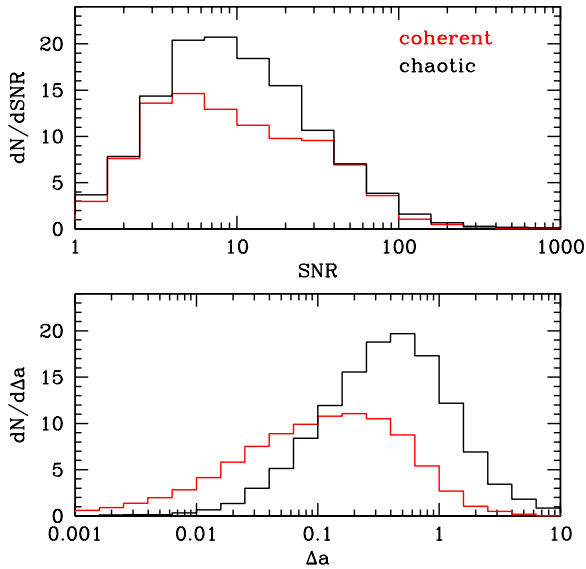}\hspace{2pc}%
\begin{minipage}[b]{18pc}\caption{\label{RDPop}SNR distribution of detected events (top histogram) and remnant spin measurement accuracy for hierarchical BH formation models with massive seeds and either coherent (red) or chaotic (black) accretion: cf.~\cite{Arun:2008zn,Berti:2008af,Sesana:2010wy} for further details. (Figure courtesy of A.~Sesana.)}
\end{minipage}
\end{figure}
The potential of a space-based mission like (e)LISA to perform
``internal'' tests of GR and constrain the merger history of massive
BHs using ringdown observations is illustrated in
Fig.~\ref{RDPop}. There we consider hypothetical (e)LISA detections of
ringdown waves (computed using analytic prescriptions from
\cite{Berti:2005ys}) within two different models for supermassive BH
formation. Both models assume a hierarchical evolution starting from
heavy BH seeds, but they differ in their prescription for the
accretion mode, which is either coherent (leading on average to large
spins) or chaotic (leading on average to small spins): see the LISA
Parameter Estimation Taskforce study \cite{Berti:2008af} for more
details.  The histograms show the distribution of SNR and spin
measurement accuracy during the two-year nominal lifetime of the eLISA
mission \cite{AmaroSeoane:2012km,AmaroSeoane:2012je}. Independently of
the accretion mode, both models predict that 1) more than ten events
would have SNR larger than 30, and 2) a few tens of events would allow
ringdown-based measurements of the remnant spin to an accuracy better
than $\sim 10\%$. Space-based detectors with six links may identify
electromagnetic counterparts to some of these merger events and
determine their distance
\cite{Arun:2008zn,AmaroSeoane:2012km,AmaroSeoane:2012je}. While
extremely promising, this simple assessment of the potential of
ringdown waves to test GR should still be viewed as somewhat
pessimistic, because a statistical ensemble of events can bring
significantly improvements over individual observations: see
e.g. \cite{Berti:2011jz} for a discussion of this point in the context
of graviton-mass bounds with (e)LISA observations of inspiralling BH
binaries\footnote{The strength of these tests will depend on two key
  elements: (i) the {\em signal-to-noise ratio} (SNR) of individual
  observations \cite{Vallisneri:2012qq}, that also affects accuracy in
  binary parameter estimation, and (ii) the {\em number $N$ of
    observations} that can be used to constrain GR.  The reason is
  that, given a theory whose deviations from GR can be parametrized by
  one or more universal parameters (e.g. coupling constants), the
  bounds on these parameters will scale roughly with $\sqrt{N}$. As a
  matter of fact, the bounds could improve faster than $\sqrt{N}$ if
  some events are particularly loud: see e.g.
  \cite{DelPozzo:2011pg,Berti:2011jz} for detailed analyses addressing
  specific modifications to GR in the Advanced LIGO/eLISA context,
  respectively.}.

\subsection{Unstable dynamics in the presence of massive bosons:
  superradiant instabilities}
\label{sec:unstable}

As anticipated at the beginning of this section, the existence of a
local minimum in the potential for massive scalar perturbations allows
for the existence of quasibound states. Detweiler
\cite{Detweiler:1980uk} computed analytically the frequencies of these
quasibound states, finding that they can induce an instability in Kerr
BHs. The physical origin of the instability is BH superradiance, as
first pointed out by Press and Teukolsky \cite{Press:1972zz} (see also
\cite{Cardoso:2004nk,Witek:2012tr,Dolan:2012yt}): scalar waves
incident on a rotating BH with frequency $0<\omega<m\Omega_H$ (where
$\Omega_H$ is the angular frequency of the horizon) extract rotational
energy from the BH and are reflected to infinity with an amplitude
which is {\em larger} than the incident amplitude. The barrier at
infinity acts as a reflecting mirror, so the wave is reflected and
amplified again. The extraction of rotational energy and the
amplification of the wave at each subsequent reflection trigger what
Press and Teukolsky called the ``black-hole bomb'' instability.

\noindent
{\bf \em Scalar fields.}
For scalar fields, results by Detweiler and others
\cite{Detweiler:1980uk,Press:1972zz,Damour:1976kh,Zouros:1979iw,Cardoso:2004nk,Dolan:2007mj,Rosa:2009ei}
show that the strenght of the instability is regulated by the
dimensionless parameter $M\mu$ (in units $G=c=1$), where $M$ is the BH
mass and $m=\mu\hbar$ is the field mass, and it is strongest when the
BH is maximally spinning and $M\mu\sim 1$
(cf.~\cite{Dolan:2007mj}). For a solar mass BH and a field of mass
$m\sim 1$~eV the parameter $M\mu\sim10^{10}\gg 1$, and the instability
is exponentially suppressed~\cite{Zouros:1979iw}. Therefore in many
cases of astrophysical interest the instability timescale must be
larger than the age of the Universe. Strong, astrophysically relevant
superradiant instabilities with $M\mu\sim1$ can occur either for light
primordial BHs which may have been produced in the early Universe,
or for ultralight exotic particles found in some extensions of the
standard model. An example is the ``string axiverse'' scenario
\cite{Arvanitaki:2009fg,Arvanitaki:2010sy}, according to which massive
scalar fields with $10^{-33}~{\rm eV}<m<10^{-18}~{\rm eV}$ could play
a key role in cosmological models. Superradiant instabilities may
allow us to probe the existence of such ultralight bosonic fields by
producing gaps in the BH Regge
plane~\cite{Arvanitaki:2009fg,Arvanitaki:2010sy} (i.e. the mass/spin
plane), by modifying the inspiral dynamics of compact
binaries~\cite{Cardoso:2011xi,Yunes:2011aa,Alsing:2011er} or by
inducing a ``bosenova'', i.e. collapse of the axion cloud (see
e.g. ~\cite{Kodama:2011zc,Yoshino:2012kn,Mocanu:2012fd}).

\begin{figure}[th]
\includegraphics[width=18pc,clip=true]{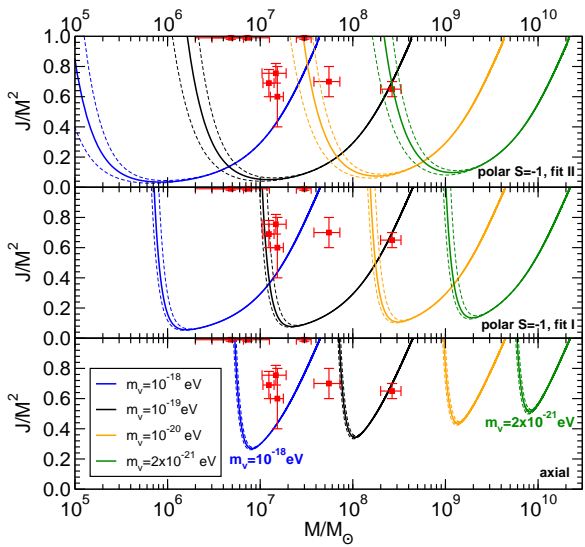}\hspace{2pc}%
\begin{minipage}[b]{18pc}\caption{\label{fig:ReggePlane}
Contour plots in the BH Regge plane~\cite{Arvanitaki:2010sy}
corresponding to an instability timescale shorter than a typical
accretion timescale, $\tau_{\rm Salpeter}=4.5\times 10^7$~yr, for
different values of the vector field mass $m_v={{\mu}}\hbar$ (from
left to right: $m_v=10^{-18}{\rm eV}$, $10^{-19}{\rm eV}$,
$10^{-20}{\rm eV}$, $2\times10^{-21}{\rm eV}$). For polar modes we
consider the $S=-1$ polarization, which provides the strongest
instability, and we use two different fits to our numerical results.
Dashed lines bracket our estimated numerical errors.  The experimental
points (with error bars) refer to the mass and spin estimates of
supermassive BHs listed in Table~2 of~\cite{Brenneman:2011wz}; the
rightmost point corresponds to the supermassive BH in
Fairall~9~\cite{Schmoll:2009gq}. Supermassive BHs lying above each of
these curves would be unstable on an observable timescale, and
therefore they exclude a whole range of Proca field masses.}
\end{minipage}
\end{figure}

\noindent
{\bf \em Vector fields.}
It has long been believed that the ``BH bomb'' instability should
operate for all bosonic field perturbations in the Kerr spacetime, and
in particular for massive spin-one (Proca)
bosons\footnote{Superradiant amplification is not possible for
  fermionic fields: see e.g.  \cite{PhysRevLett.31.1265,Iyer:1978du}.}
\cite{Dolan:2007mj}. A proof of this conjecture was lacking until
recently because of technical difficulties in separating the
perturbation equations for massive spin-one (Proca) fields in the Kerr
background. Pani et al. recently circumvented the problem using a
slow-rotation expansion pushed to second order in rotation
\cite{Pani:2012vp,Pani:2012bp}.  The Proca superradiant instability
turns out to be stronger than the massive scalar field
instability. Furthermore the Proca mass range where the instability
would be active is very interesting from an experimental point of
view: indeed, as shown in \cite{Pani:2012vp}, astrophysical BH spin
measurements are already setting the most stringent upper bound on the
mass of spin-one fields.  This can be seen in
Fig.~\ref{fig:ReggePlane}, which shows exclusion regions in the ``BH
Regge plane'' (cf. Fig.~3 of~\cite{Arvanitaki:2010sy}) obtained by
setting the instability timescale equal to the (Salpeter) accretion
timescale $\tau_{\rm Salpeter}=4.5\times 10^7$~yr. The idea here is
that a conservative bound on the critical mass of the Proca field
corresponds to the case where the instability spins BHs down {\em
  faster} than accretion could possibly spin them up. Instability
windows are shown for four different masses of the Proca field
($m_v=10^{-18}$~eV, $10^{-19}$~eV, $10^{-20}$~eV and
$2\times10^{-21}$~eV) and for two different classes of unstable Proca
modes: ``axial'' modes (bottom panel) and ``polar'' modes with
polarization index $S=-1$, which provides the strongest instability
(top and middle panels). All regions above the instability window are
ruled out\footnote{While our numerical results for the axial modes are
  supported by an analytical formula, in the polar case we have used
  two different functions to fit the numerical data at second order in
  the BH spin.}. The plot shows that essentially \emph{any} spin
measurement for supermassive BHs with $10^6M_\odot\lesssim M\lesssim
10^9M_\odot$ would exclude a wide range of vector field masses
\cite{Pani:2012vp,Pani:2012bp}. Massive vector instabilities do not --
strictly speaking -- provide ``external'' tests of GR, but rather
tests of perturbative dynamics within GR; quite interestingly, they
provide constraints on possible mechanisms to generate massive
``hidden'' U(1) vector fields, which are predicted by various
extensions of the Standard Model
\cite{Goldhaber:2008xy,Goodsell:2009xc,Jaeckel:2010ni,Camara:2011jg}. The
results discussed in this section are quite remarkable, because they
show that astrophysical measurements of nonzero spins for supermassive
BHs can already place the strongest constraints on the mass of
hypothetical vector bosons (for comparison, the Particle Data Group
quotes an upper limit $m<10^{-18}~$eV on the mass of the photon
\cite{Beringer:1900zz}).

\section{Present and future tests of massive scalar-tensor theories}
\label{sec:GWs}

So far I discussed ``internal'' tests of GR from future GW
observations of stable BH dynamics (ringdown waves). I also summarized
how superradiant instabilities can be used to place bounds on the
masses of scalar and vector fields, which emerge quite naturally in
extensions of the Standard Model
\cite{Arvanitaki:2009fg,Arvanitaki:2010sy,Goldhaber:2008xy,Goodsell:2009xc,Jaeckel:2010ni,Camara:2011jg}. In
this Section I address a slightly different but related question,
namely: what constraints on the mass and coupling of scalar fields are
imposed by Solar System observations? Shall we be able to constrain
these models better (or prove that scalar fields are indeed needed for
a correct description of gravity) using future GW observations?

\subsection{Solar System bounds}

In \cite{Alsing:2011er} we investigated observational bounds on
massive scalar-tensor theories of the Brans-Dicke type. In addition to
deriving the orbital period derivative due to gravitational radiation,
we also revisited the calculations of the Shapiro time delay and of
the Nordtvedt effect in these theories (cf. \cite{Will:2005va} for a
detailed and updated treatment of these tests).

\begin{figure}[th]
\includegraphics[width=18pc,clip=true]{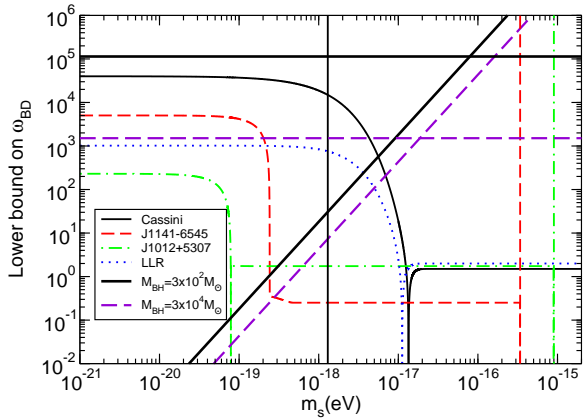}\hspace{2pc}%
\begin{minipage}[b]{18pc}\caption{\label{fig:bounds_all}
Lower bound on $(\omega_{\rm BD}+3/2)$ as a function of the mass of
the scalar $m_s$ from the Cassini mission data (black solid line;
cf. \cite{Perivolaropoulos:2009ak}), period derivative observations of
PSR J1141-6545 (dashed red line) and PSR J1012+5307 (dot-dashed green
line), and Lunar Laser Ranging experiments (dotted blue
line). Vertical lines indicate the masses corresponding to the typical
radii of the systems: 1AU (black solid line) and the orbital radii of
the two binaries (dashed red and dot-dashed green lines).  Note that
the theoretical bound on the coupling parameter is $\omega_{\rm
  BD}>-3/2$.}
\end{minipage}
\end{figure}

The comparison of our results for the orbital period derivative,
Shapiro time delay and Nordtvedt parameter against recent
observational data allows us to put constraints on the parameters of
the theory: the scalar mass $m_s$ and the Brans-Dicke coupling
parameter $\omega_{\rm BD}$. These bounds are summarized in Figure
\ref{fig:bounds_all}. We find that the most stringent bounds come from
the observations of the Shapiro time delay in the Solar System
provided by the Cassini mission (which had already been studied in
\cite{Perivolaropoulos:2009ak}). From the Cassini observations we
obtain $\omega_{\rm BD}>40,000$ for
$m_s<2.5\times10^{-20}\mathrm{eV}$, while observations of the
Nordtvedt effect using the Lunar Laser Ranging (LLR) experiment yield
a slightly weaker bound of $\omega_{\rm BD}>1,000$ for
$m_s<2.5\times10^{-20}\mathrm{eV}$.  Possibly our most interesting
result concerns observations of the orbital period derivative of the
circular white-dwarf neutron-star binary system PSR J1012+5307, which
yield $\omega_{\rm BD}>1,250$ for $m_s<10^{-20}\mathrm{eV}$. The
limiting factor here is our ability to obtain precise measurements of
the masses of the component stars as well as of the orbital period
derivative, once kinematic corrections have been accounted
for. However, there is considerably more promise in the eccentric
binary PSR J1141-6545, a system for which remarkably precise
measurements of the orbital period derivative, the component star
masses and the periastron shift are available.  The calculation in
\cite{Alsing:2011er} was limited to circular binaries, and we are
currently working to generalize our treatment to eccentric binaries in
order to carry out a more meaningful and precise comparison with
observations of PSR J1141-6545.

\subsection{Gravitational-wave tests}

Binary pulsar observations can test certain aspects of strong-field
modifications to GR, such as the ``spontaneous scalarization''
phenomenon in scalar-tensor theories \cite{Damour:1996ke}, and
interesting tests are also possible with current astronomical
observations \cite{Psaltis:2008bb}. However a real breakthrough is
expected to occur in the near future with the {\em direct} detection
of GWs from the merger of compact binaries composed of BHs and/or NSs.
One of the most exciting prospects of the future network of GW
detectors (Advanced LIGO/Virgo \cite{2010CQGra..27h4006H}, LIGO-India
\cite{indigo} and KAGRA \cite{Somiya:2011np} in the near future;
third-generation Earth-based interferometers like the Einstein
Telescope \cite{2010CQGra..27s4002P} and a space-based, LISA-like
mission \cite{Danzmann:1998,AmaroSeoane:2012km,AmaroSeoane:2012je} in
the long term) is precisely their potential to test GR in
strong-field, high-velocity regimes inaccessible to Solar System and
binary pulsar experiments.
Second-generation interferometers such as Advanced LIGO should detect
a large number of compact binary coalescence events
\cite{Abadie:2010cf,Dominik:2012kk}. Unfortunately from the point of
view of testing GR, most binary mergers detected by Advanced
LIGO/Virgo are expected to have low signal-to-noise ratios (a possible
exception being the observation of intermediate-mass BH mergers
\cite{AmaroSeoane:2009ui}, that would be a great discovery in and by
itself). Third-generation detectors such as the Einstein Telescope
will perform significantly better in terms of parameter estimation and
tests of alternative theories \cite{Sathyaprakash:2012jk,Gair:2010dx}.
Here I will argue (using the example of massive scalar-tensor
theories) that an (e)LISA-like mission will be an ideal instrument to
test GR \cite{AmaroSeoane:2012km,AmaroSeoane:2012je} by providing two
examples: (1) bounds on massive scalar-tensor theories using (e)LISA
observations of intermediate mass-ratio inspirals, and (2) the
possibility to observe an exotic phenomenon related once again to
superradiance, i.e., floating orbits.

\noindent
{\bf \em Bounds on massive scalar-tensor theories from intermediate
  mass-ratio inspirals.}
%
In general, the gravitational radiation from a binary in massive
scalar-tensor theories depends on both the scalar field mass $m_s$ and
the coupling constant $\omega_{\rm BD}$
\cite{Berti:2004bd,Alsing:2011er}. If the field is massless,
corrections to the GW phasing are proportional to $1/\omega_{\rm BD}$,
and therefore comparisons of the phasing in GR and in scalar-tensor
theories yield bounds on $\omega_{\rm BD}$
\cite{Will:1994fb,Berti:2004bd}. By computing the GW phase in the
stationary-phase approximation, one finds that the scalar mass always
contributes to the phase in the combination $m_s^2/\omega_{\rm BD}$,
so that GW observations of nonspinning, quasicircular inspirals can
only set upper limits on $m_s/\sqrt{\omega_{\rm BD}}$
\cite{Berti:2012bp}. For large SNR $\rho$, the constraint is inversely
proportional to $\rho$.  The order of magnitude of the achievable
bounds is essentially set by the lowest frequency accessible to the GW
detector, and it can be understood by noting that the scalar mass and
GW frequency are related (on dimensional grounds) by $m_s({\rm
  eV})=6.6\times 10^{-16}\, f({\rm Hz})$, or equivalently $f({\rm
  Hz})=1.5\times 10^{15}\, m_s({\rm eV})$.
%
For eLISA, the lower cutoff frequency (imposed by acceleration noise)
$f_{\rm cut}\sim 10^{-5}$~Hz corresponds to a scalar of mass
$m_s\simeq 6.6\times 10^{-21}$~eV. For Earth-based detectors the
typical seismic cutoff frequency is $f_{\rm cut}\sim 10$~Hz,
corresponding to $m_s\sim 6.6\times 10^{-15}$~eV. This simple
argument shows that space-based detectors can set $\sim 10^6$ stronger
bounds on the scalar mass than Earth-based detectors.

An explicit calculation shows that the best bounds are obtained from
(e)LISA observations of the intermediate mass-ratio inspiral of a
neutron star into a BH of mass $M_{\rm BH}\lesssim 10^3~M_\odot$, and
that they would be of the order
\be 
\left(\frac{m_s}{\sqrt{\omega_{\rm BD}}}\right)
\left(\frac{\rho}{10}\right)\lesssim 10^{-19}~{\rm eV}.
\ee
In summary, GW observations will provide two constraints: a lower
limit on $\omega_{\rm BD}$ (corresponding to horizontal lines in
Fig.~\ref{fig:bounds_all}) and an upper limit on
$m_s/\sqrt{\omega_{\rm BD}}$ (corresponding to the straight diagonal
lines in Fig.~\ref{fig:bounds_all}). Therefore GW observations would
exclude the complement of a trapezoidal region on the top left of
Fig.~\ref{fig:bounds_all}.  Straight (dashed) lines show the bounds
from eLISA observations of NS-BH binaries with SNR $\rho=10$ when the
BH has mass $M_{\rm BH}=300~M_\odot$ ($M_{\rm BH}=3\times
10^4~M_\odot$, respectively). The plot shows that GW observations with
$\rho=10$ become competitive with binary pulsar bounds when
$m_s\gtrsim 10^{-19}$~eV, and competitive with Cassini bounds when
$m_s\gtrsim 10^{-18}$~eV, with the exact ``transition point''
depending on the SNR of the observation (for a GW observation with SNR
$\rho=100$ the ``straight line'' bounds in Fig.~\ref{fig:bounds_all}
would be ten times higher). Therefore in this particular theory a
single high-SNR observation (or the statistical combination of several
observations, see e.g. \cite{Berti:2011jz}) may yield better bounds on
the scalar coupling than weak-gravity observations in the Solar System
when $m_s\gtrsim 10^{-18}$~eV.

\noindent
{\bf \em Floating orbits.}
It is generally expected that small bodies orbiting around a BH will
lose energy in gravitational waves, slowly inspiralling into the
BH. In \cite{Cardoso:2011xi} we showed that the coupling of a massive
scalar field to matter leads to a surprising effect: because of
superradiance, orbiting objects can hover into ``floating orbits'' for
which the net gravitational energy loss at infinity is entirely
provided by the BH's rotational energy. The idea is that a compact
object around a rotating BH can excite superradiant modes to
appreciable amplitudes when the frequency of the orbit matches the
frequency of the unstable quasibound state. This follows from energy
balance: if the orbital energy of the particle is $E_p$, and the total
(gravitational plus scalar) energy flux is $\dot E_T= \dot E^{g}+\dot
E^{s}$, then
\begin{equation}
\dot E_p+\dot E^{g}+\dot E^{s}=0\,.\label{balance}
\end{equation}
Usually $\dot E^{g}+\dot E^{s}>0$, and therefore the orbit shrinks
with time. However it is possible that, due to superradiance, $\dot
E^{g}+\dot E^{s}=0$. In this case $\dot E_p=0$, and the orbiting body
can ``float'' rather than spiralling in
\cite{Misner:1972kx,Press:1972zz}. The system is essentially a ``BH
laser'', where the orbiting compact object is producing stimulated
emission of {\em gravitational} radiation: because the massive scalar
field acts as a mirror, negative scalar radiation ($\dot E^{s}<0$) is
dumped into the horizon, while gravitational radiation can be detected
at infinity.  Orbiting bodies remain floating until they extract
sufficient angular momentum from the BH, or until perturbations or
nonlinear effects disrupt the orbit. For slowly rotating and
nonrotating BHs floating orbits are unlikely to exist, but resonances
at orbital frequencies corresponding to quasibound states of the
scalar field can speed up the inspiral, so that the orbiting body
``sinks''. A detector like (e)LISA could easily observe these effects
\cite{Cardoso:2011xi,Yunes:2011aa}, that would be spectacular smoking
guns of deviations from general relativity.

\section{Conclusions}

The three examples discussed in this paper (ringdown tests of the BH
no-hair theorem, bosonic superradiant instabilities in rotating BHs
and GW tests of massive scalar-tensor theories) illustrate that
astrophysical BHs, either in isolation or in compact binaries, can be
spectacular nature-given laboratories for fundamental physics. We can
already use astrophysical observations to do fundamental physics
(e.g. by setting bounds on the masses of scalar and vector fields
using supermassive BH spin measurements), but the real goldmine for
the future of ``fundamental astrophysics'' will be GW observations. In
order to fully realize the promise of GWs as probes of strong-field
gravity we will need several detections with large SNR. Second- and
third-generation Earth-based interferometers will certainly deliver
interesting science, but a full realization of strong-field tests and
fundamental physics with GW observations may have to wait for
space-based GW detectors. We'd better make sure they happen in our
lifetime.

\vspace{.5cm}

%
%


\begin{acknowledgements}
The research reviewed in this paper was supported by NSF CAREER Grant
No. PHY-1055103. I thank my collaborators on various aspects of the
work described in this paper: Justin Alsing, Vitor Cardoso, Sayan
Chakrabarti, Jonathan Gair, Leonardo Gualtieri, Michael Horbatsch,
Akihiro Ishibashi, Paolo Pani, Alberto Sesana, Ulrich Sperhake, Marta
Volonteri, Clifford Will and Helmut Zaglauer. Special thanks go to
Paolo Pani for comments on an early draft and to Alberto Sesana for
preparing Fig.~\ref{RDPop}, as well as excellent mojitos.
\end{acknowledgements}

\bibliographystyle{spphys}       


\end{document}